\documentclass[aps,titlepage,12pt,superscriptaddress]{revtex4}
\usepackage{epsfig}
\usepackage{times}
\usepackage{color}
\usepackage{cases}
\usepackage{amssymb}
\begin{document}
\title{Braess's Paradox in Epidemic Game: Better Condition Results in Less Payoff}

\author{Hai-Feng Zhang}
\affiliation{School of Mathematical Science, Anhui University, Hefei 230039, P. R. China}

\author{Zimo Yang}
\affiliation{Web Sciences Center, University of Electronic Science and Technology of
China, Chengdu 611731, P. R. China}

\author{Zhi-Xi Wu}
\affiliation{Institute of Computational Physics and Complex Systems,
Lanzhou University, Lanzhou 730000, P. R. China}
\author{Bing-Hong Wang}
\affiliation{Department of Modern Physics, University of Science and
Technology of China, Hefei 230026, P. R. China}
\author{Tao Zhou}
\email{zhutou@ustc.edu}
\affiliation{Web Sciences Center, University of Electronic Science and Technology of
China, Chengdu 611731, P. R. China}

\begin{abstract}
Facing the threats of infectious diseases, we take various
actions to protect ourselves, but few studies considered
an evolving system with competing strategies. In view of
that, we propose an evolutionary epidemic model coupled with human behaviors, where individuals have three strategies: vaccination,
self-protection and laissez faire, and could adjust their strategies according to their neighbors' strategies
and payoffs at the beginning of each new season of epidemic spreading. We found a counter-intuitive
phenomenon analogous to the well-known \emph{Braess's Paradox}, namely a better condition may lead to worse
performance. Specifically speaking, increasing the successful rate
of self-protection does not necessarily reduce the epidemic size
or improve the system payoff. This phenomenon is insensitive to the
network topologies, and can be well explained by a mean-field
approximation. Our study demonstrates an important fact that a better
condition for individuals may yield a worse outcome for the society.

\end{abstract}

\maketitle

\section{Introduction}
Recent outbreaks of global infectious diseases, including SARS
(Severe Acute Respiratory Syndrome), H1N1 (Swine Influenza) and H5H1
(Avian Influenza), have caused major public healthy threats owing to
their potential mortalities and substantial economic impacts.
According to the report of WHO, infectious diseases cause
more than 10 million deaths annually and accounting for 23\% of the
global disease burden~\cite{world2004world}. Various interventions
thus have been developed to control infectious diseases, such as
vaccination, treatment, quarantining and behavior change programs
(e.g., social distancing and partner
reduction)~\cite{enns2011optimal}.

Though preemptive vaccination is the fundamental method for
preventing transmission of infectious diseases as well as reducing
morbidity and
mortality~\cite{bauch2003group,bauch2004vaccination,perisic2009simulation},
practically, the immunization of individuals is more than a
voluntary behavior owing to the economic costs, logistical
limitations, religious reasons, side effects, and so
on~\cite{schimit2011vaccination}. Therefore, instead of vaccinating,
people may prefer to take some self-protective actions including
reducing outside activities, detouring to avoid epidemic areas,
wearing face masks, washing hands frequently, and so
forth~\cite{meloni2011modeling,perra2011towards,fenichel2011adaptive,sahneh2012existence}.
Generally speaking, these self-protective actions are less costly
and cannot guarantee the safety against the diseases.

Under such complicated environment, an individual's strategy usually
results from a tradeoff between cost and risk. For instance, people
may be laissez-faire to the spreading of common flu, while they will
take vaccination for hepatitis B since the vaccines are very
effective and hepatitis B is very difficult to be eradicated. In
contrast, people prefer to take self-protection against HIV since
its consequence is terrible while the effectivity and side effects
of vaccines are both unknown. Accordingly, game-theoretic models may be
suitable to characterize these decision-making
processes~\cite{bauch2003group,bauch2004vaccination,bauch2005imitation,vardavas2007can,basu2008integrating,perisic2009social}.
Bauch \emph{et al.}~\cite{bauch2003group,bauch2004vaccination}
analyzed population behavior under voluntary vaccination policies
for childhood diseases via a game-theoretic framework, and they
found that voluntary vaccination is unlikely to reach the
group-level optimum due to the risk perception in vaccines and the
effect of herd immunity. Bauch~\cite{bauch2005imitation} studied a game model in which individuals adopt strategies according to an imitation
dynamics, and found that oscillations in vaccine uptake can emerge under different conditions, for example, vaccinating behavior is very sensitive to the changes in disease prevalence. Vardavas \emph{et al.}~\cite{vardavas2007can} considered the effects of voluntary
vaccination on the prevalence of influenza based on a minority game, and found that severe epidemics could not be prevented unless
vaccination programs offer incentives. Basu \emph{et
al.}~\cite{basu2008integrating} proposed an epidemic game model for
HPV vaccination based on the survey data on actual perceptions
regarding cervical caner, showing that the actual vaccination level
is far lower than the overall vaccination goals. Perisic and
Bauch~\cite{perisic2009social} studied the interplay between
epidemic spreading dynamics and individual vaccinating behavior on
social contact networks. Compared with the homogeneously mixing
model, they found that increasing the neighborhood size of the
contact network can eliminate the disease if individuals decide
whether to vaccinate by accounting for infection risks from neighbors.

As mentioned above, in most related works, individuals are usually
divided into two opposite classes: vaccinated and laissez-faire,
while less attention is paid on other alterative strategies in
between. In this paper, we propose an evolutionary epidemic game
model to study the effects of self-protection on the system payoff
and epidemic size. We find a counter-intuitive phenomenon analogous to the well-known \emph{Braess's Paradox} \cite{Braess1968} in network traffic, that is, the
increasing of successful rate of self-protection may, on the
contrary, decrease the system payoff. We provide a mean-field solution, which
well reproduces such observations. This study raises an unprecedent
challenge on how to lead the masses of people when facing an
epidemic, since sufficient knowledge about and effective protecting
skills to the infectious disease, which sound very helpful for every
individual, may eventually enlarge the epidemic size and cause
losses for the society.

\section{Model definition}
Considering a seasonal flu-like disease that spreads through a
social contact network~\cite{fu2011imitation,wu2011imperfect}. At
the beginning of a season, each individual could choose one of the
three strategies: vaccination, self-protection or laissez faire. If
an individual gets infected during this epidemic season, she will pay
a cost $r$. A vaccinated individual will pay a cost $c$ that
accounts for not only the monetary cost of the vaccine, but also the
perceived vaccine risks, side effects, long-term healthy impacts,
and so on. We assume that the vaccine could perfectly protect
vaccinated individuals. A self-protective individual will pay a less
cost $b$, while a laissez-faire individual pays nothing. Denote
$\delta$ the successful rate of self-protection, that is, a
self-protective individual will be equivalent to a vaccinated
individual with probability $\delta$ or be equivalent to a
laissez-faire individual with probability $1-\delta$. This will be determined right after an individual's decision for simplicity. Obviously,
$r>c>b>0$. Without loss of generality, we set the cost of being
infected as $r=1$. Table 1 presents the payoffs for different
strategies and outcomes.

When the strategy of every individual is fixed, all individuals can
be divided into two classes: susceptible ones including
laissez-faire individuals and part of self-protective individuals,
and irrelevant individuals (they are equivalent to be removed from the system) including vaccinated ones and all other
self-protective individuals. The susceptible individuals probably
get infected while the irrelevant individuals will not affect or be
affected by the epidemic dynamics. Among all susceptible
individuals, $I_{0}$ individuals are randomly selected and set to be
infected initially. The spreading dynamics follows the standard
susceptible-infected-removed (SIR)
model~\cite{anderson1992infectious}, where at each time step, each
infected individual will infect all her susceptible neighbors with
probability $\lambda$, and then she will turn to be a removed
individual with probability $\mu$. The spreading ends when no
infected individual exists. Then, the number of removed individuals,
$R^\infty$, is called the epidemic size or the prevalence.

\begin{table}[tbp]
\centering \caption{\label{tab:0} The payoffs for different
strategies and outcomes.}
\begin{tabular}{lccc}  
\hline
 &   Healthy & Infected  \\ \hline  
Laissez-faire &     $0$&   $-1$\\         
Self-protected &     $-b$&  $-1-b$\\        
 Vaccinated&    $-c$&  N/A\\ \hline
\end{tabular}
\end{table}

After this epidemic season, every individual updates her strategy by
imitating her neighborhood. Firstly, she will randomly select one
neighbor and then decide whether to take this neighbor's strategy.
We apply the Fermi
rule~\cite{traulsen2006stochastic,perc2010coevolutionary}, namely an
individual $i$ will adopt the selected neighbor $j$'s strategy with
probability
\begin{equation}\label{1}
W(s_i\leftarrow s_j)=\frac{1}{1+\exp[-\kappa(P_j-P_i)]},
\end{equation}
where $s_i$ means the strategy of $i$, $P_i$ is $i$'s payoff in the last season, and the parameter $\kappa>0$
characterizes the strength of selection: smaller $\kappa$ means that
individuals are less responsive to payoff difference. After the moment
all individuals have decided their strategies (and thus their roles
in the epidemic spreading are also decided), a new season starts.

\section{Results}
We first study the model on square lattices with von Neumann neighborhood and periodic boundary conditions. Figure~\ref{fig1}(a) presents the effects of the successful rate of self-protection, $\delta$, on the decision makings of individuals and the epidemic size. Clearly, as the increasing of $\delta$, the condition gets better and better. A counter-intuitive phenomenon is observed when $\delta$ lies in the middle range (from about 0.3 to about 0.4), during which a better condition leads to a larger epidemic size. One may think that though the epidemic size becomes larger, the system payoff (the sum payoff of all individuals) could still get higher since individuals pay less in choosing self-protection than vaccination. However, as shown in figure~\ref{fig1}(b) and ~\ref{fig1}(c), the system payoff is strongly negatively correlated with the epidemic size. That is to say, a better condition (i.e., a larger $\delta$) could result in worse performance in view of both the larger epidemic size and the less system payoff. This is very similar to the so-called \emph{Braess's Paradox}, which states that adding extra capacity to a network when the moving entities selfishly choose their route, can in some cases reduce overall performance \cite{Braess1968,Roughgarden2005}.

Figure~\ref{fig2} shows the strategy distribution patterns of four representative cases. When $\delta$ is small, it is unwise to take self-protection because of its low efficiency, and people prefer to take vaccination or laissez faire. As shown in figure~\ref{fig2}(b), there are only two strategies, vaccination and laissez faire, and thus $\delta$ has no effect on the epidemic size. Both infected and not infected laissez-faire individuals form percolating clusters, and are nearly (not fully) separated by vaccinated individuals. Of course, this kind of partial separation can only be possible when the number of vaccinated individuals is considerable. When $\delta$ gets larger, more and more individuals take self-protection and fewer and fewer individuals take vaccination or laissez faire. Since only a fraction, $\delta$, of self-protective individuals are equivalent to the vaccinated individuals, the system contains more susceptible individuals. In addition, these susceptible individuals are less protected since the number of irrelevant individuals becomes smaller. Such two factors lead to the increase of the epidemic size and the decrease of the system payoff. As shown in figure~\ref{fig2}(c), the light-red percolating cluster is fragmented into pieces due to the decrease of irrelevant individuals, which is also a reason of the decrease of the fraction of laissez-faire individuals: being laissez-faire becomes more risky now. When $\delta$ is large, the superiority of self-protection becomes more striking and no one takes vaccination, then the epidemic size decreases as $\delta$ increases. As shown in figure~\ref{fig2}(d), self-protective and laissez-faire individuals coexist. As the increasing of $\delta$, though the  self-protection strategy is more efficient, the laissez faire strategy is more attractive since the irrelevant individuals becomes more and thus for susceptible individuals, the risk of being infected becomes smaller. This is the reason why the fraction of laissez-faire individuals become more and more in the right range. In fact, when $\delta$ is very large, the laissez-faire and not infected individuals again form a percolating cluster. Please see figure~\ref{fig2}(e) for the example case at $\delta=0.95$.

Figures S1-S5 verify the universality of the counter-intuitive phenomenon. Figure S1 reports the epidemic size as a function of $\delta$ for square lattices with different sizes, suggesting that our main results are insensitive to the network size. To verify the insensitivity to network structures, we implement the model on disparate networks including the Erd\"os-R\'enyi (ER) networks~\cite{erdds1959random}, the Barab\'asi-Albert (BA) networks~\cite{barabasi1999emergence} and the well-mixed networks (i.e., fully connected networks or called complete networks). As shown in figure~\ref{fig3}, in despite to the quantitative difference, the counter-intuitive phenomenon is observed for all kinds of networks. Figures S2-S5 present systematical simulation results about the effects of different parameters on different kinds of networks. For every kind of networks, one can observe the counter-intuitive phenomenon when the condition $0<b<c<1$ is hold.

Although the phenomenon is qualitatively universal for different kinds of networks, as shown in figure~\ref{fig3}, there are quantitative differences between square lattices and other kinds of networks: (i) in ER, BA and well-mixed networks, the self-protection strategy gets promoted and could become the sole strategy in a certain range of $\delta$; (ii) the epidemic size in ER, BA and well-mixed networks is smaller than that in square lattices. In square lattices, laissez-faire individuals could form clusters that are guarded by the surrounding irrelevant individuals. Then they paid nothing but can escape from the infection. On the contrary, ER, BA and well-mixed networks do not display localized property and thus to choose laissez-faire strategy is of high risk. Therefore, with delocalization, the laissez-faire strategy is depressed while the self-protection strategy gets promoted and less individuals will get infected.

To verify the above inference, we remove a number of edges in the square lattice and randomly add the same number of edges. During this randomizing process, the network connectivity is always guaranteed and the self-connections and multi-connections are always not allowed. The number of removed edges, $A$, can be used to quantify the strength of delocalization. As shown in figure~\ref{fig4}, with the increasing of $A$, the self-protection strategy gets promoted and the clusters of not infected laissez-faire individuals are fragmented into small pieces. When $A$ gets larger and larger, the strategy distribution pattern becomes closer and closer to that of ER, BA and well-mixed networks. The gradually changing process in figure 4 clearly demonstrates that the main reason resulting in the quantitative differences is the structural localization effects. In a word, the ER, BA and well-mixed display essentially the same results since they do not have many localized clusters.

Lastly, we present an analytical solution based on the mean-field approximation for well-mixed networks (see \textbf{Methods}), which could reproduce the counter-intuitive phenomenon. Figure~\ref{fig5} compares the analytical prediction with simulation, indicating a good accordance.

\section{Discussion}
Spontaneous behavioral responses to epidemic situation are recognized to have
significant impacts on epidemic spreading, and thus to incorporating human behavior into epidemiological models can
enhance the models' utility in mimicking the reality and evaluating control measures \cite{zhang2012modeling,zhang2010hub,salathe2008effect,poletti2011effect,funk2009spread,coelho2009dynamic,funk2010modelling}. To this end, we proposed an evolutionary epidemic game where individuals can choose their strategies towards infectious diseases and adjust their strategies according to their neighbors' strategies and payoffs.

Strikingly, we found a counter-intuitive phenomenon that a better condition (i.e., larger successful rate of self-protection) may unfortunately result in less system payoff. It is because when the successful rate of self-protection increases, people become more speculative and less interested in vaccination. Since a vaccinated individual indeed brings benefit to the system by statistically reducing the infection probability of susceptible individuals, the decreasing of vaccinated individuals will eventually lead to the loss of system payoff. Qualitatively speaking, the counter-intuitive phenomenon is insensitive to the network topology, while quantitatively speaking, networks with delocalized structure (e.g., ER, BA and well-mixed networks) have more self-protective individuals and less laissez-faire individuals than networks with localized structure (e.g., square lattices), and the epidemic size is larger in the latter case. Without the diverse behavioral responses of individuals, epidemic in delocalized structure usually spreads more quickly and widely than in localized structure \cite{Eguiluz2002,Zhou2006}. The opposite observation reported in the current model again results from more and more speculative choices (i.e., to be laissez-faire) at a low-risky situation. Therefore, this can be considered as another kind of ``less payoff in better condition" phenomenon.

The observed counter-intuitive phenomenon reminds us of the well-known Braess's Paradox in network traffic \cite{Braess1968,Roughgarden2005}. Zhang \emph{et al.} \cite{Zhang2007} showed that to remove some specific edges in a network can largely enhance its information throughput, and Youn \emph{et al.} \cite{Youn2008} pointed out that some roads in Boston, New York City and London could be closed to reduce predicted travel times. Actually, Seoul has removed a highway to build up a park, which, beyond all expectations, maintained the same traffic but reduced the travel time \cite{Baker2009}. Very recently, Pala \emph{et al.} \cite{Pala2012} showed that Braess's Paradox may occur in mesoscopic electron systems, that is, adding a path for electrons in a nanoscopic network may paradoxically reduce its conductance. This work provides another interesting example analogous to Braess's Paradox, namely a higher successful rate of self-protection may eventually enlarge the epidemic size and thus cause system loss. Let's think of the prisoner's dilemma, if every prisoner stays silent, they will be fine, while one more choice, to betray, makes the situation worse for them. Analogously, if the successful rate $\delta$ is small, few people will choose to be self-protective, while for larger $\delta$, people have more choices, which may eventually reduce the number of vaccinated people and thus enlarge the epidemic size. Basically, both the original Braess's Paradox and the current counter-intuitive phenomenon are partially due to the additional choices to selfish individuals. This is easy to be understood in a simple model like the prisoner's dilemma game, but it is impressive to observe such phenomenon in a complex epidemic game.

Human-activated systems are usually much more complex than our expectation, since people's choices and actions are influenced by the environment and at the same time their choices and actions have changed the environment. This kind of interplay leads to many unexpected collective responses to both emergencies and carefully designed policies, which, fortunately, can still be modeled and analyzed to some extent. This work raises an unprecedent challenge to the public health agencies about how to lead the population towards an epidemic. The government should take careful consideration on how to distribute their resources and money on popularizing vaccine, hospitalization, self-protection, self-treatment, and so on.

\section{Methods}
Given a well-mixed network with size $N$, the dynamical equations are
\begin{eqnarray}
  \frac{dS}{dt} &=& -\lambda N SI, \label{1a}\\
  \frac{dI}{dt} &=& \lambda N SI-\mu I,\label{1b} \\
  \frac{dR}{dt} &=&\mu I,\label{1c}
\end{eqnarray}
where $S$, $I$ and $R$ stand for the fraction of susceptible, infected and recovered individuals, respectively. Dividing Eq. (2) by Eq. (4), one has
\begin{equation}
\frac{dS}{dR}=-R_0S,
\end{equation}
where $R_0=\frac{\lambda N}{\mu}$ is the basic reproduction number for the standard SIR model in well-mixed population~\cite{anderson1992infectious}. Integrating Eq. (5), we get
\begin{equation}
\int_{S(0)}^{S(\infty)}\frac{dS}{S}=\int_{R(0)}^{R(\infty)}-R_0dR,
\end{equation}
which leads to the solution
\begin{equation}
\ln \frac{S(\infty)}{S(0)}=-R_0[R(\infty)-R(0)].
\end{equation}
Clearly, $R(0)=0$, $R(\infty)+S(\infty)=1$, and in the thermodynamic limit, $S(0)\approx 1$. Accordingly,
\begin{equation}\label{2}
    R(\infty)=1-\exp\left[-R_{0}R(\infty)\right].
\end{equation}

Let $\rho_V$, $\rho_S$ and $\rho_L$ be the fraction of vaccinated, self-protective and laissez-faire individuals, such that $\rho_V+\rho_S+\rho_L\equiv1$. Since only a fraction $1-\rho_V+\delta \rho_S$ of individuals are susceptible, using the similar techniques, one can easily obtain the epidemic size as
\begin{equation}
    R'(\infty)=(1-\rho_V-\delta \rho_S)\left\{1-\exp\left[-R'_0R'(\infty)\right]\right\},
\end{equation}
where $R'_0=(1-\rho_V-\delta \rho_S)R_0$. Then, the probability of a susceptible individual to be infected reads
\begin{equation}
\omega=\frac{R'(\infty)}{1-\rho_V-\delta \rho_S}=1-\exp\left[-R'_0R'(\infty)\right].
\end{equation}
The payoffs of different strategies and states are thus easily to be obtained, which are summarized in Table 2.

\begin{table}[tbp]
\centering \caption{\label{tab:2} The payoffs for different
strategies and states. $P_V$, $P_S$ and $P_L$ stand for average payoffs for individuals with strategy vaccination, self-protection and laissez faire, while the superscripts $H$ (healthy) and $I$ (infected) represent the final states. }
\begin{tabular}{cccc}  
\hline
\hline
Strategy \& State &   Fraction & Payoff  \\ \hline   

Vaccinated \& Healthy &     $\rho_V$&   $P_V=-c$\\          

Self-protective \&
Healthy&$\rho_S^H=\rho_S[\delta+(1-\delta)(1-\omega)]$& $P_S^H=-b$\\        

Self-protective \&  Infected&
$\rho_S^I=\rho_S(1-\delta)\omega$&
$P_S^I=-b-1$\\

 Laissez-faire \&  Healthy&    $\rho_L^H=(1-\rho_V-\rho_S)(1-\omega)$&  $P_L^H=0$\\
Laissez-faire \& Infected&
$\rho_L^I=(1-\rho_V-\rho_S)\omega$&
$P_L^I=-1$\\
\hline
\hline
\end{tabular}
\end{table}

The imitation dynamics governing the time evolution of the fractions of strategies
in the population is similar to the replicator dynamics of evolutionary game
theory~\cite{wu2011imperfect,poletti2009spontaneous}, as
\begin{eqnarray}
  \frac{d\rho_V}{dt} &&= (\rho_V\rightleftarrows \rho_{S}^H)+(\rho_V\rightleftarrows \rho_{S}^I)
  +(\rho_V\rightleftarrows \rho_{L}^H)+(\rho_V\rightleftarrows \rho_{L}^I),\label{5}\\
  \nonumber\frac{d\rho_S}{dt} &&= (\rho_{S}^H\rightleftarrows \rho_V)+(\rho_{S}^H\rightleftarrows \rho_{L}^H)
  +(\rho_{S}^H\rightleftarrows \rho_{L}^I)+(\rho_{S}^I\rightleftarrows \rho_V)
  \\&&+(\rho_{S}^I\rightleftarrows \rho_{L}^H)+(\rho_{S}^I\rightleftarrows \rho_{L}^I)\label{6},
\end{eqnarray}
where
\begin{eqnarray}
\nonumber  \rho_V\rightleftarrows
\rho_{S}^H&=&(\rho_{S}^H\rightarrow \rho_V)-(\rho_V\rightarrow
\rho_{S}^H) \\
\nonumber  &=&\rho_V\rho_{S}^H\left\{\frac{1}{1+\exp\left[-\kappa(P_V-P_S^H)\right]}-\frac{1}{1+\exp\left[-\kappa(P_S^H-P_V)\right]}\right\} \\
\nonumber  &=&\rho_V\rho_{S}^H\tanh\left[\frac{\kappa}{2}(P_V-P_S^H)\right] \\
&=&\rho_V\rho_S[\delta+(1-\delta)(1-\omega)]\tanh\left[\frac{\kappa}{2}(-c+b)\right],
\end{eqnarray}
and the others are similar.

Denote by $\rho_V(\tau)$ the initial fraction of vaccinated individuals before the $(\tau+1)$th season of epidemic spreading. Given $\rho_V(0)$, $\rho_S(0)$ and $\rho_L(0)$, and for each season, we apply the initial conditions as $S(0)=(N'-5)/N'$, $I(0)=5/N'$ and $R(0)=0$, where $N'=(1-\rho_V-\delta \rho_S)N$, depending on the distribution of strategies at this season. Then, $R'(\infty)$ can be obtained by Eq. (9) and $\omega$ by Eq. (10). Using the evolutionary dynamics described in Eqs. (11)-(13) and the fractions presented in Table 2, one can obtain the values of $\rho_V(1)$, $\rho_S(1)$ and $\rho_L(1)$, which are also the initial fractions of strategies at the beginning of the next season. Repeat the above steps until the steady state, then we can calculate the desired variables.

\begin{acknowledgments}
This work was partially supported by the National Natural Science
Foundation of China under Grant Nos. 11005001, 11005051, 11222543, 11135001, 11275186, 91024026 and 10975126.  H.-F.Z. acknowledges the Doctoral Research Foundation of Anhui University under Grant No. 02303319. T.Z. acknowledges the Program for New Century Excellent Talents in University under Grant No. NCET-11-0070.

\end{acknowledgments}

\clearpage

\begin{figure}
\centerline{\epsfig{file=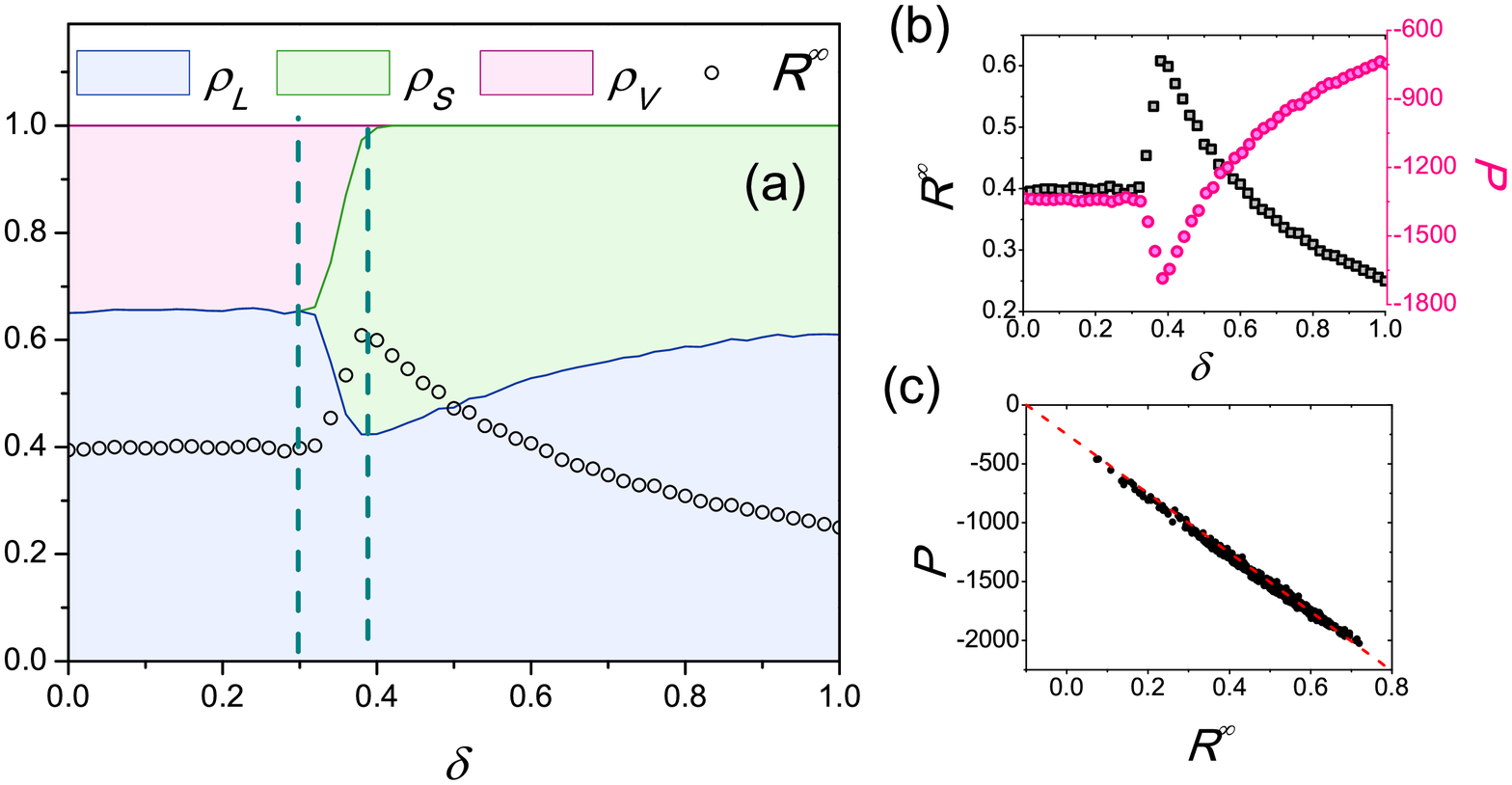,width=16cm}}
\caption{\label{fig1} \textbf{Less payoff in better condition}. (a) How the fractions of the three strategies and the epidemic size change with the successful rate of self-protection $\delta$. (b) The epidemic size $R^{\infty}$ and the system payoff $P$ as functions of $\delta$. (c) Correlation between the system payoff $P$ and the epidemic size $R^{\infty}$, where each data point corresponds to a certain $\delta$. Panel (a) is divided into three regions by two vertical dash lines: (i) In the left region, no self-protective individual exists and $\delta$ has no effect on the epidemic size; (ii) In the middle region, the self-protection strategy gradually replaces vaccination and laissez faire, and the epidemic size increases with $\delta$ due to the decrement of vaccination fraction; (iii) In the right region, with high successful rate of self-protection, individuals are unwilling to take vaccination and the epidemic size decreases with $\delta$.  Parameters are set to be $N=50\times50=2500$, $\lambda=0.5$, $\mu=0.3$, $b=0.1$, $c=0.4$, $\kappa=10$ and $I_{0}=5$. For this figure and all others (except snapshots), the simulation results are calculated after 1000 seasons when the system is in a steady state, and each data point is obtained by averaging over 100 independent runs.}\label{fig1}
\end{figure}

\begin{figure}
\centerline{\epsfig{file=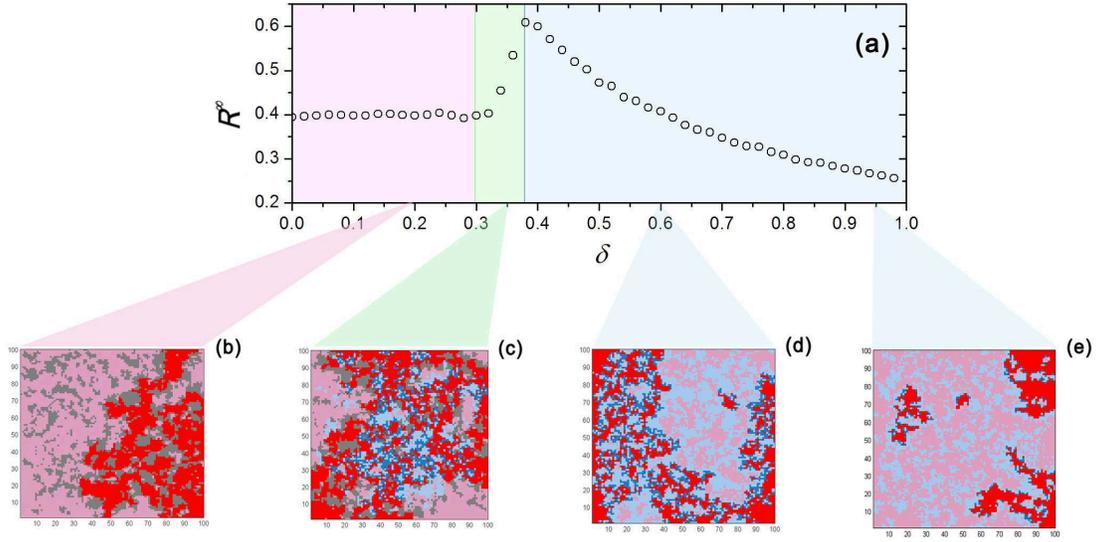,width=16cm}}
\caption{\label{fig2}\textbf{Strategy distribution patterns}. Subgraph (a) shows the epidemic size $R^{\infty}$ as a function of $\delta$. The window is divided into three parts according to the tendency of $R^{\infty}-\delta$ curve. Subgraphs (b), (c), (d) and (e) are snapshots in the steady state of a season at $\delta=0.2$, 0.35, 0.5 and 0.95. The grey, light red, dark red, light blue and dark blue points stand for vaccinated, laissez-faire and not infected, laissez-faire and infected, self-protective and not infected, and self-protective and infected individuals, respectively. Parameters are the same as in Figure~\ref{fig1}.}
\end{figure}

\begin{figure}
\epsfig{file=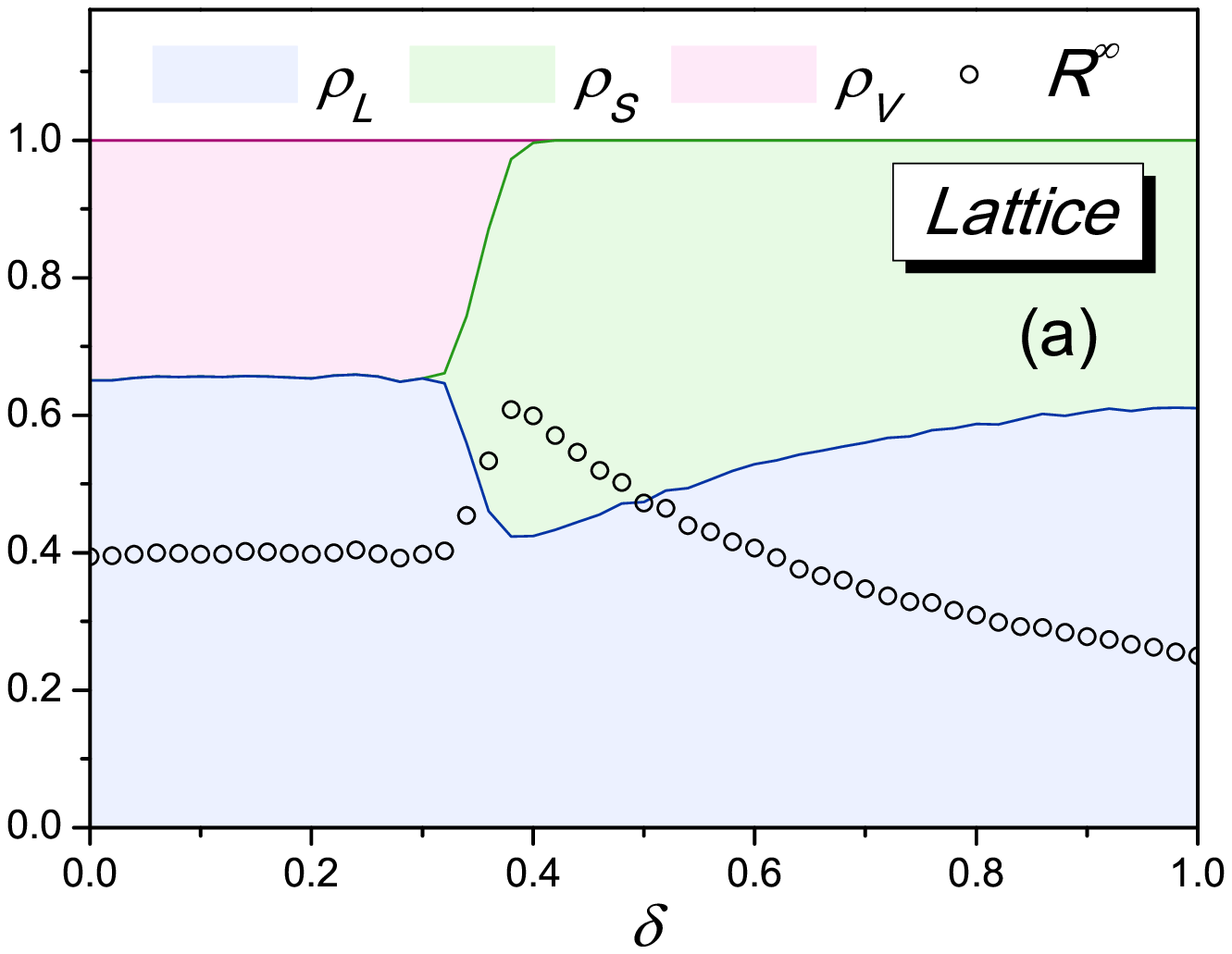,width=8cm}
\epsfig{file=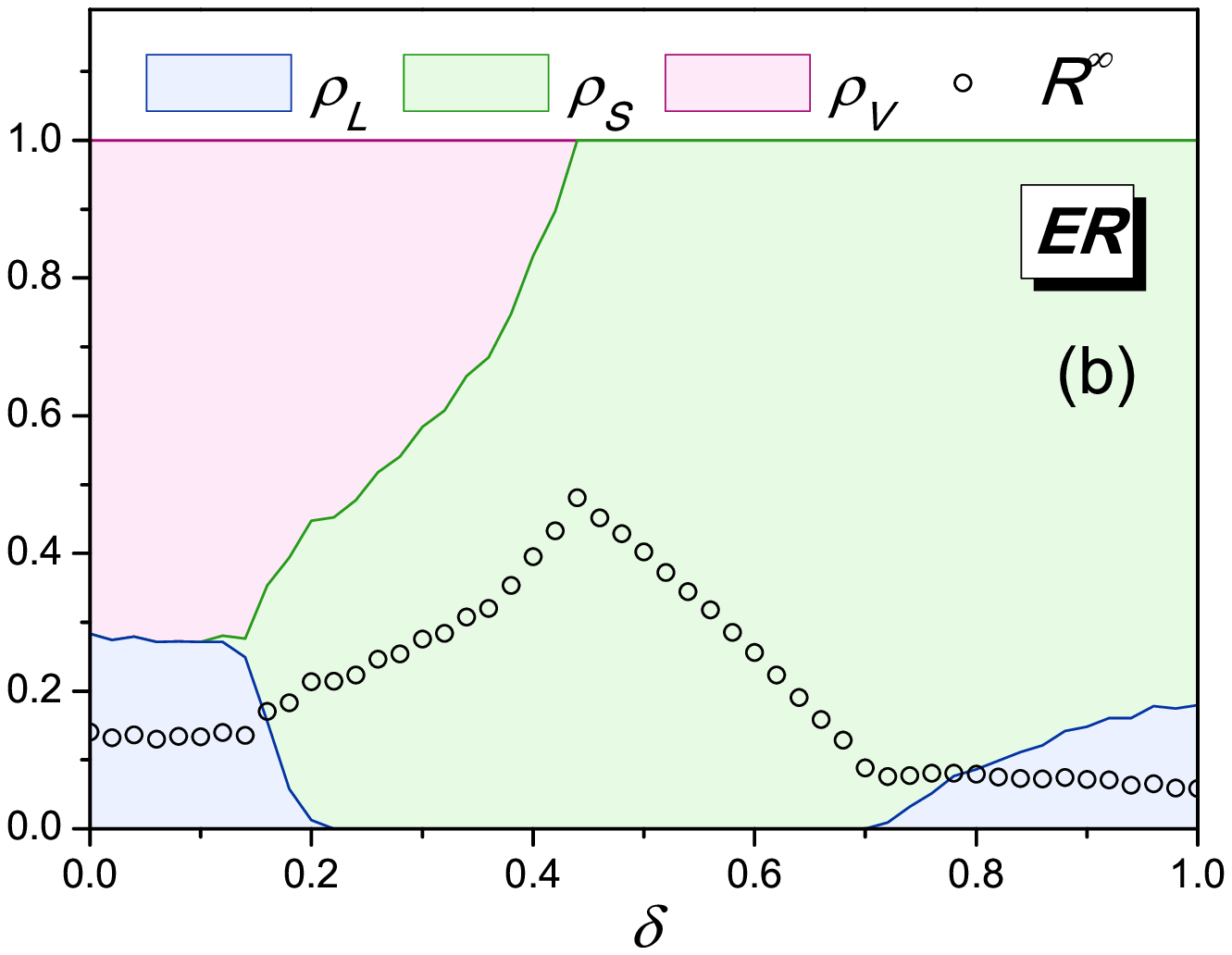,width=8cm}
\epsfig{file=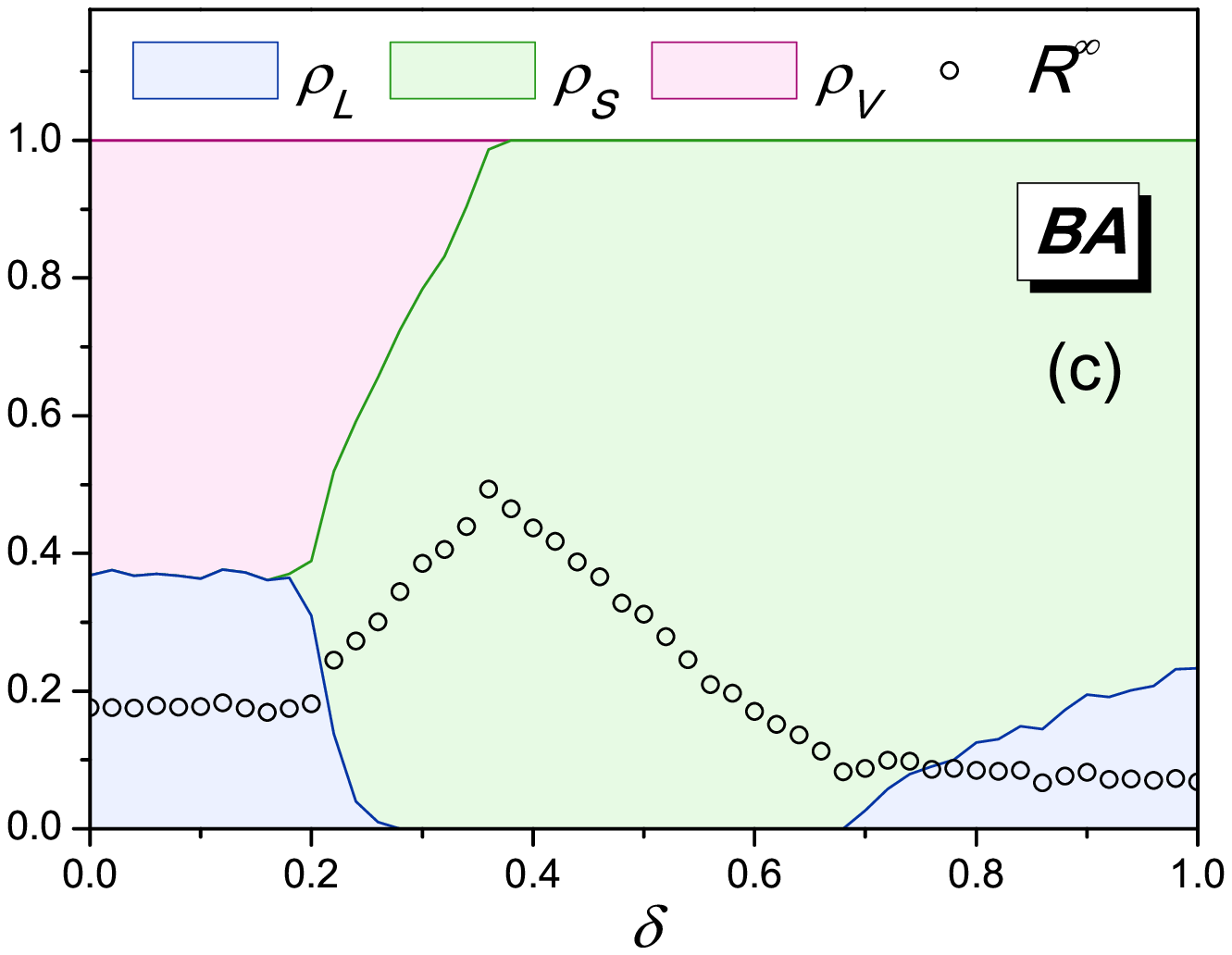,width=8cm}
\epsfig{file=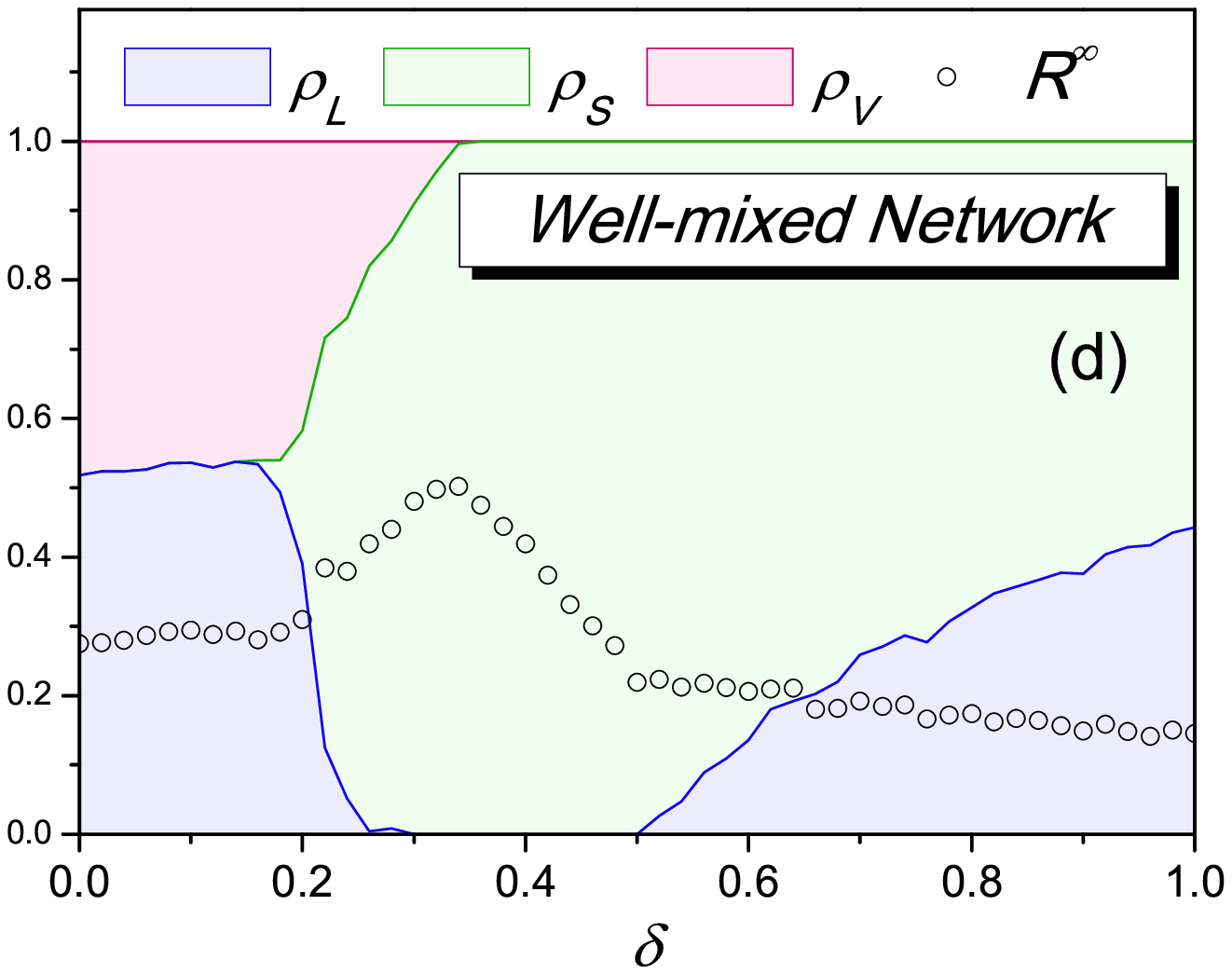,width=8cm}
\caption{\label{fig3} \textbf{Insensitivity to the network structures}. To explore the impacts of different network structures on the epidemic size and strategy distribution, we compare the results in square lattices (a), ER networks (b), BA networks (c) and well-mixed networks (d). The parameters are set as $b=0.1$, $I_0=5$, $c=0.4$, and $\kappa=10$. Each data point results from an average over 100 independent runs. The average degrees of the lattices, ER networks and BA networks are all set to be 4, and the simulations presented in subgraph (a), (b) and (c) are implemented with the same transmission and recovery rate, $\lambda=0.5$ and $\mu=0.3$. For the well-mixed network (d), however, the parameters are different from others as $\lambda=0.0013$ and $\mu=1$ for its different average degree.}
\end{figure}

\begin{figure}
\centerline{\epsfig{file=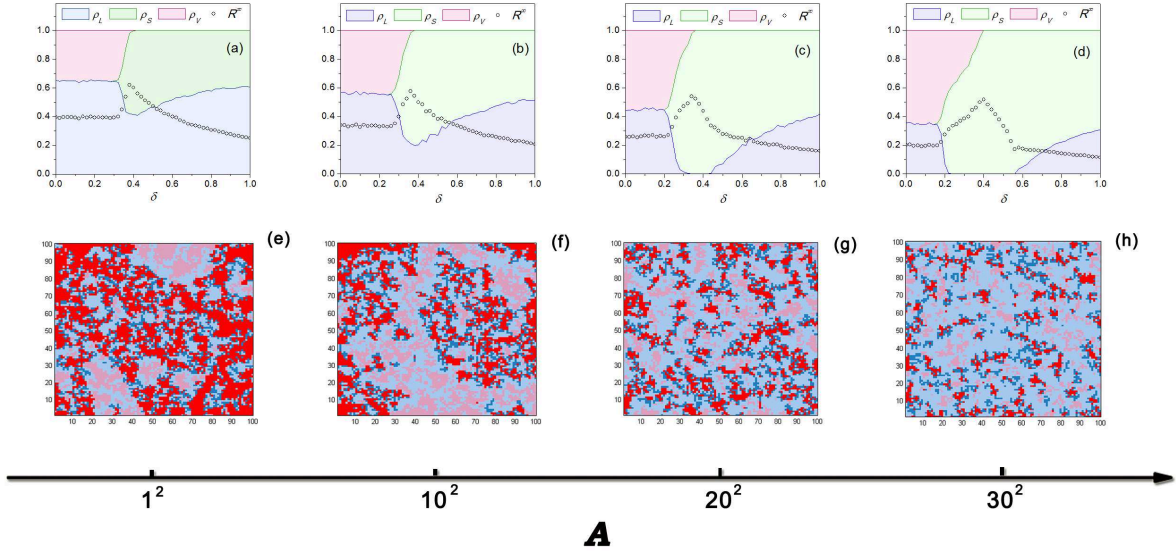,width=16cm}}
\caption{\label{fig4}\textbf{Delocalization promotes the self-protection strategy}. The subgraphs (a)-(d) show how the fractions of the three strategies and the epidemic size change with $\delta$, and subgraphs (e)-(h) are the corresponding snapshots for (a)-(d) with $\delta=0.6$. From (a) to (d), the number of randomized edges, $A$, increases. Qualitatively, the counter-intuitive phenomenon always exists, no matter what the value of $A$. Quantitatively, the delocalization reduces the advantage of the laissez-faire strategy, which leads to a larger fraction of self-protective individuals. When $A$ is large enough, self-protection becomes the dominating strategy for a certain range of $\delta$. Overall speaking, the epidemic size is smaller at larger $A$. Parameters are the same as in Figure~\ref{fig1}.}
\end{figure}

\begin{figure}
\centerline{\epsfig{file=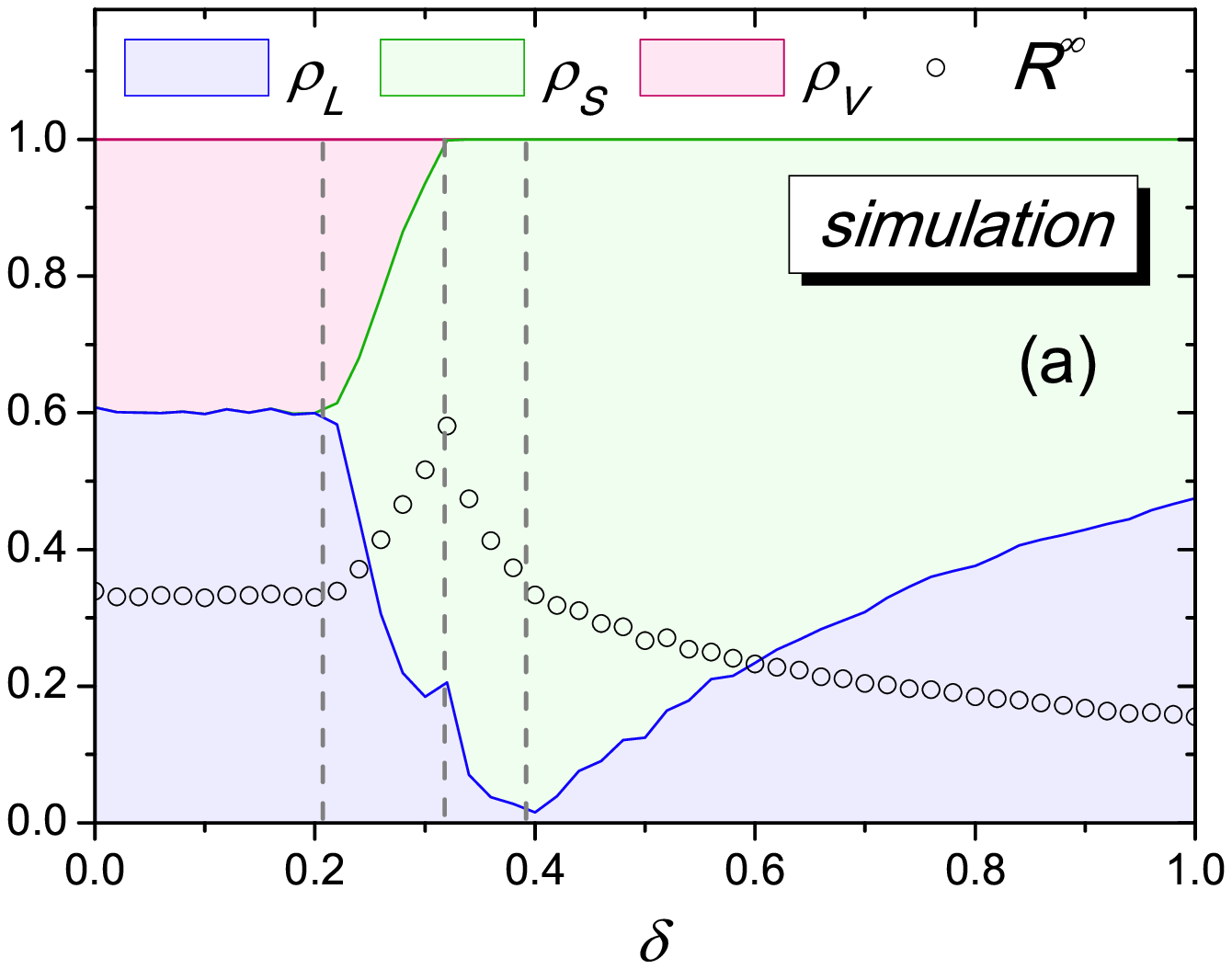,width=8cm}\epsfig{file=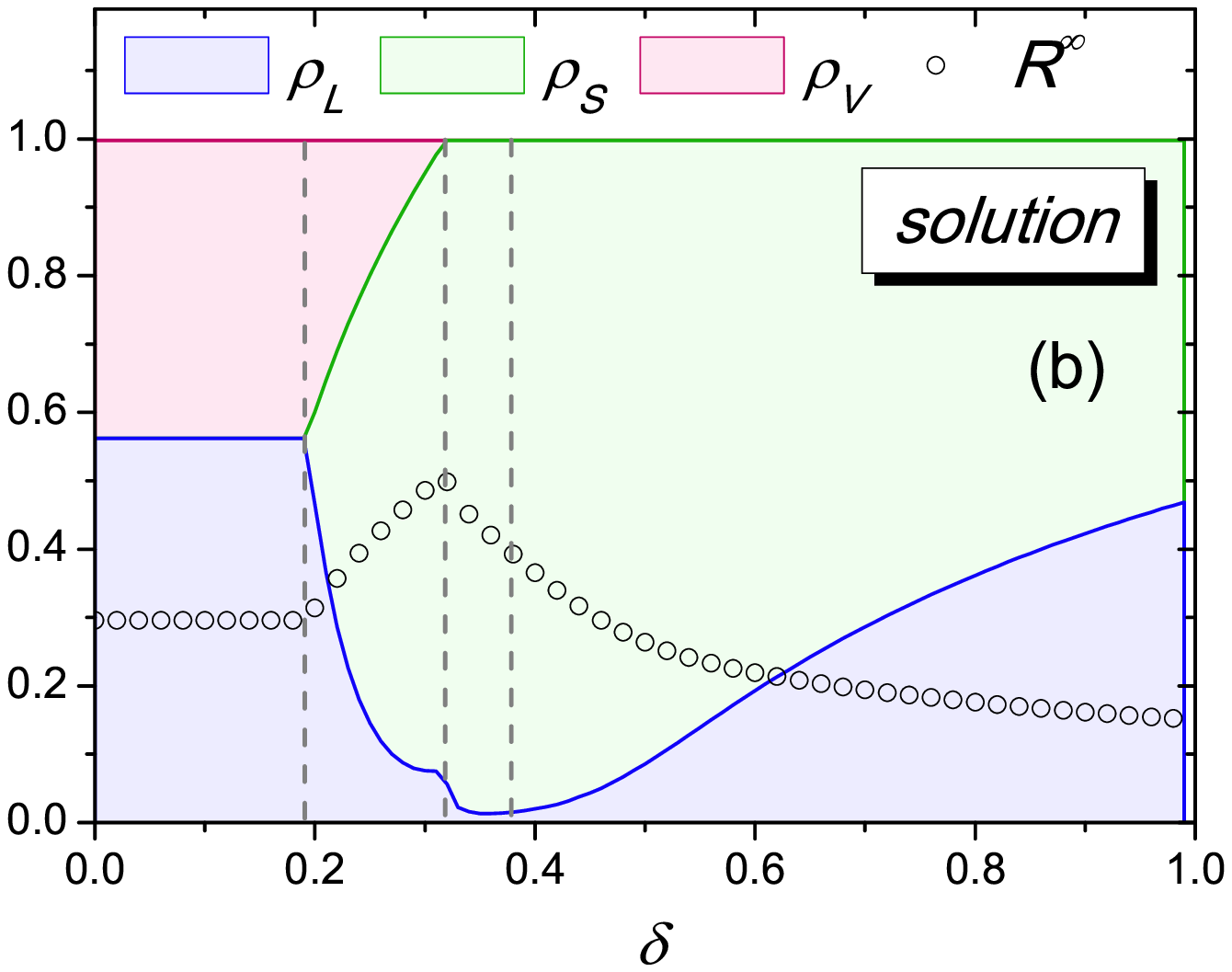,width=8cm}}
\caption{\label{fig5}\textbf{The analytical solution agrees well with the simulation}.
The analytical prediction (b) is in good accordance with the simulation (a). All results are implemented on a well-mixed network with $N=1000$, $c=0.7$, $b=0.1$, $\lambda=0.0013$, $\mu=1.0$, $I_0=5$ and $\kappa=10$.}
\end{figure}

\end{document}